\documentstyle[flushrt,psfig]{slides}
\begin{document}

\title{The Equilibrium Structure of Cosmological Halos: From Dwarf
Galaxies to X-ray Clusters}

\author{Paul R. Shapiro and Ilian T. Iliev,\\
The University of Texas at Austin
}
\date{13 March 2000}
\maketitle
{\footnotesize
\centerline{\underline{Abstract}$^*$}

A new model for the postcollapse equilibrium structure of virialized objects
 which condense out of the cosmological background universe is described 
and compared with observations and simulations of cosmological
halos from dwarf galaxies to X-ray clusters. The model is based upon the 
assumption that virialized halos are isothermal, which leads to a prediction
of a unique nonsingular isothermal sphere for the equilibrium structure, 
with a core radius which is approximately 1/30 times the size and a core 
density which is proportional to the mean background density at the epoch 
of collapse.
These predicted nonsingular isothermal spheres are in good agreement with the
observations of the internal structure of dark-matter-dominated halos from 
dwarf galaxies to X-ray clusters. 

\vspace{2cm}

$^*$ Poster paper presented at the Institute for Theoretical Physics Conference 
on Galaxy Formation and Evolution, March 14-17, 2000, at the
University of California at Santa Barbara.

\pagebreak

\centerline{\underline{Equilibrium structure of virialized halos}}
[Shapiro, Iliev, and Raga 1999, MNRAS, 307, 203 (Einstein-de Sitter case);
Iliev and Shapiro 2000 (low-density, open and flat with cosmological constant)] 
\begin{itemize}
\item Problem and Motivation 
\begin{itemize}
\item Question: What equilibrium structure forms when a density 
perturbation collapses out of the expanding background universe and virializes?
\item An analytical model for the structure (e.g. mass profile, temperature, velocity
dispersion, radius) of virialized cosmological halos would be a valuable tool
for the semi-analytical modeling of galaxy and cluster formation in a hierarchical
clustering model like CDM.
\begin{itemize}
\item Earlier work adopted crude approximations involving either uniform spheres or
singular isothermal spheres which resulted from top-hat perturbation collapse
\item What is a more realistic outcome, even for the simple top-hat problem?
\end{itemize}
\item N-body simulations of CDM predict dark matter halo profiles with singular density
profiles, but a finite density core is required to explain: 
\begin{itemize}
\item Dwarf galaxy rotation curves
\item Cluster mass profiles inferred from gravitational lensing
\end{itemize}
\item As a result, the cold, collisionless nature of CDM has recently been re-examined
to allow for variations which affect the post-collapse equilibrium structure of halos.
\item Suppose we ignore the details of this relaxation process and adopt the assumption 
that the final equilibrium is isothermal.
\item Solve this problem and compare the result with dwarf galaxy rotation curves
and X-ray cluster data. 
\end{itemize}
\item Model: 
\begin{itemize}
\item Top-hat density perturbation collapses and virializes 
\item Virialization leads to a truncated isothermal sphere in 
hydrostatic equilibrium (TIS) $\Rightarrow$ solution of the Lane-Emden 
equation (modified for $\Lambda\neq 0$)
\item Total energy of top-hat is conserved thru collapse and virialization
\item Postcollapse temperature set by virial theorem (including effect of
 finite boundary pressure)
\end{itemize}
%\pagebreak
\item Is the solution uniquely determined? -- No, some additional information 
is required:
\begin{itemize}
\item[1)] Minimum-Energy Solution: Boundary pressure is that for which the 
conserved top-hat energy is the minimum possible for an isothermal sphere 
of fixed mass within a finite truncation radius.
\item[2)] The Self-Similar Spherical Cosmological Infall Solution (Bertschinger 1985)
confirms this choice if we identify the virialized object with the spherical 
region of post-shock gas and shell-crossing dark matter $\Rightarrow$ explains
dynamical origin of boundary pressure adopted above as the result of thermalizing 
the energy of infall.
\end{itemize}
\pagebreak
{\footnotesize
\centerline{\underline{Summary of the TIS Solution}}
\vskip-1cm
\item Top-hat perturbation $\Rightarrow$ unique, {\bf nonsingular} TIS
(minimum-energy configuration) 

$\qquad\Rightarrow$ universal, 
self-similar density profile for the  postcollapse equilibrium of cosmic 
structure
\begin{itemize}
\item Unique scale and amplitude set by top-hat mass and collapse epoch
\item Same density profile for gas and dark matter (no bias)
\end{itemize}
\item[I.]Matter-Dominated Case (see Table 1 and Fig. 1)
\begin{itemize}
\item Finite core size: $r_0= 0.034\,\,\times$ radius $r_t$
\item Central density: $\rho_0= 514 \,\,\times$ surface density $\rho_t$
\item $T=2.16\,\, T_{\rm uniform\,\, sphere}
	=0.72\,\, T_{\rm singular\,\,isothermal\,\, sphere}$
\item At intermediate radii, $\rho$ drops faster than $r^{-2}$
\end{itemize}}
\item[II.] Flat, $\Lambda\neq0$ Case
\begin{itemize}
\item Profile varies with epoch of collapse, approaching case
 I above for early collapse.

For example: for $\Omega_0=1-\lambda_0=0.3$ for $z_{coll}=(0;0.5;1)$:
\begin{itemize}
\item $r_t/r_0=(30.04;29.68;29.54)$ 
\item $\rho_0/\rho_t=(529.9;520.8;517.2)$ 
\item $T/T_{\rm uniform\,\, sphere}=(2.188;2.170;2.163)$
\end{itemize}
\end{itemize}
\end{itemize}
\pagebreak
\centerline{\underline{Table 1: The Postcollapse Virial Equilibrium Resulting}}

\centerline{\underline{from the Collapse of Top-Hat Density Perturbations:}}

\centerline{\underline{Einstein-de Sitter Universe}}

\begin{minipage}{250mm}
\begin{tabular}{@{}lcccc}
&Uniform&Singular& Our\\
& Sphere&Isothermal&Solution$^*$\\
&& Sphere&\\\hline
$\eta=\frac{r_t}{r_m}$.....&0.5&0.417&0.554\\[7mm]
$\displaystyle{\frac{k_BT_{\rm vir}}
	{\left(\frac25\frac{GMm}{r_{\rm vir}}\right)}}$...&1&3&2.16\\[15mm]
$\displaystyle{\frac{\rho_0}{\rho_t}}$...........& 1&$\infty$&514\\[7mm]
$\displaystyle{\frac{\langle\rho\rangle}{\rho_t}}$...........&1&3&3.73\\[7mm]
$\displaystyle{\frac{r_t}{r_0}}..............$& -- NA --&$\infty$&29.4\\[7mm]
$\displaystyle{\frac{\langle\rho\rangle}{\rho_{b}(t_{\rm coll})}}$.....&$18\pi^2$&
      $\displaystyle{18\pi^2\left(\frac65\right)^3\approx\pi^5}$&\\[5mm]
&$\,\,\,\,\,\,\,\approx 178$
	&$\approx 307$&130.5
\\[5mm]\hline
\end{tabular}
\end{minipage}
\vskip0.5cm
{$^*$ Our solution = minimum-energy, truncated,}
{ nonsingular, isothermal sphere}

Note: $\rho_b\equiv$ cosmic mean matter density
\pagebreak

\centerline{\underline{Density Profile of Halo which Forms from Top-Hat}}

\centerline{\underline{ Perturbation Collapse and Virialization}}
  
\centerline{\psfig{figure=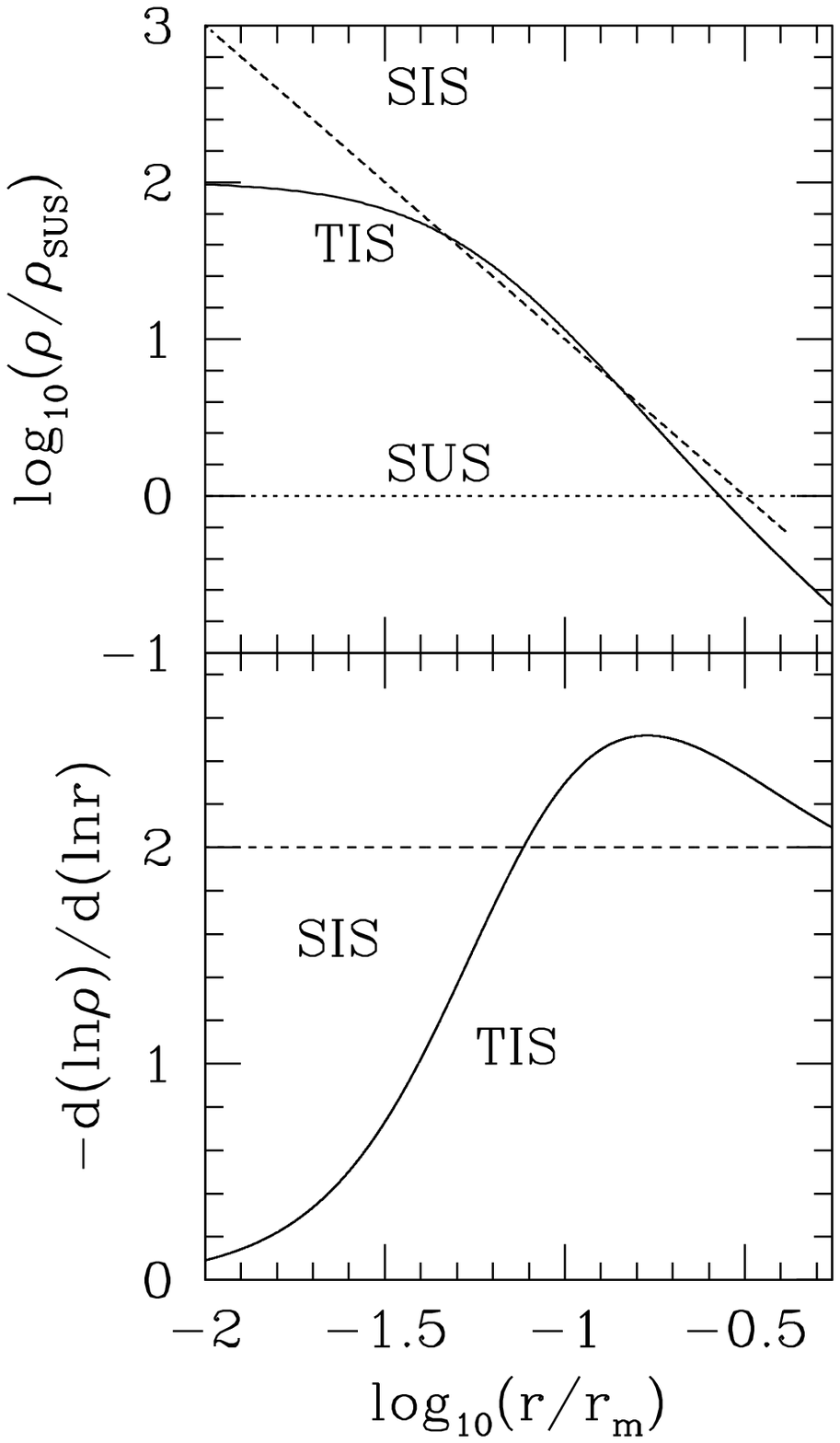,height=6.5in,width=6.5in}}
%\vspace{-1cm}

Fig.1: Density profile of truncated isothermal sphere which forms from the virialization
of a top-hat density perturbation in a matter-dominated universe. Radius $r$ is in
units of $r_m$ - the top-hat radius at maximum expansion, while density $\rho$ is
in terms of the density $\rho_{SUS}$ of the standard uniform sphere approximation for 
the virialized, post-collapse top-hat. (TIS = our solution, SUS = uniform sphere,
SIS = singular isothermal sphere). Bottom panel shows logarithmic slope of density 
profile.

\pagebreak

\centerline{\underline{Direct Comparison with NFW Profile}}
\centerline{\psfig{figure=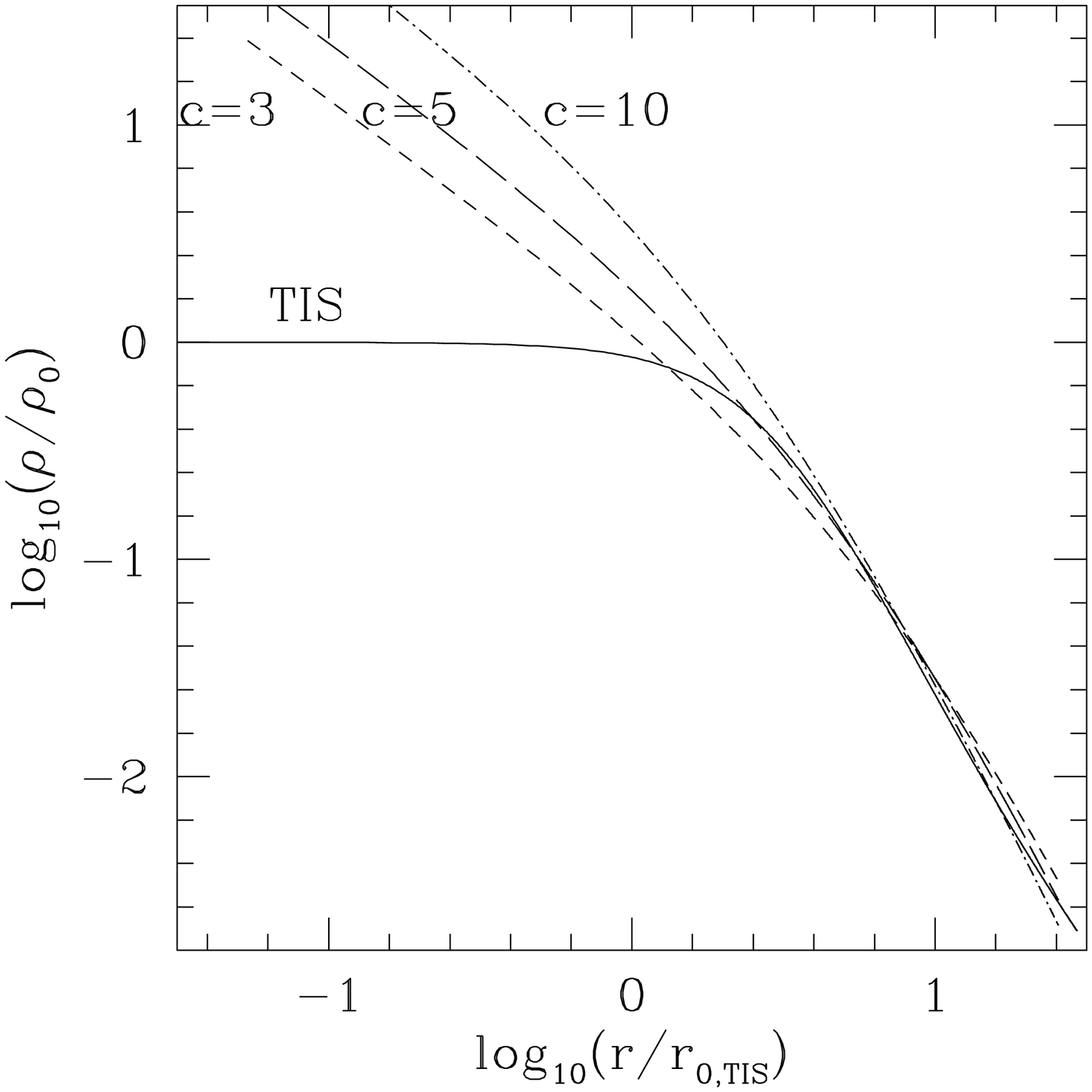,height=6.5in,width=6.5in}}
%\vspace{-1cm}

Fig. 2: Continuous line = TIS profile;
Dashed lines = ``NFW'' = Navarro, Frenk, and White (1996, 1997) profile:
$$
\rho(r)=\frac{\delta_c\rho_{b0}}{cx(cx+1)^2},\qquad x=\frac r{r_{200}}
$$
 Range of $c$ appropriate for X-ray clusters to early forming dwarf galaxies.
\pagebreak

\centerline{\underline{Dwarf Galaxy Rotation Curves}} 

Q: How well does our TIS profile match the observed mass profiles of 
dark-matter-dominated dwarf galaxies? The observed rotation curves of
dwarf galaxies can be fit according to the following density profile
with a finite density core (Burkert 1995):
$$
\rho(r)=\frac{\rho_{0,Burkert}}{(r/r_c+1)(r^2/r_c^2+1)}
$$

A: The TIS profile gives a nearly perfect fit to the Burkert profile. (see Fig. 3)

\centerline{\psfig{figure=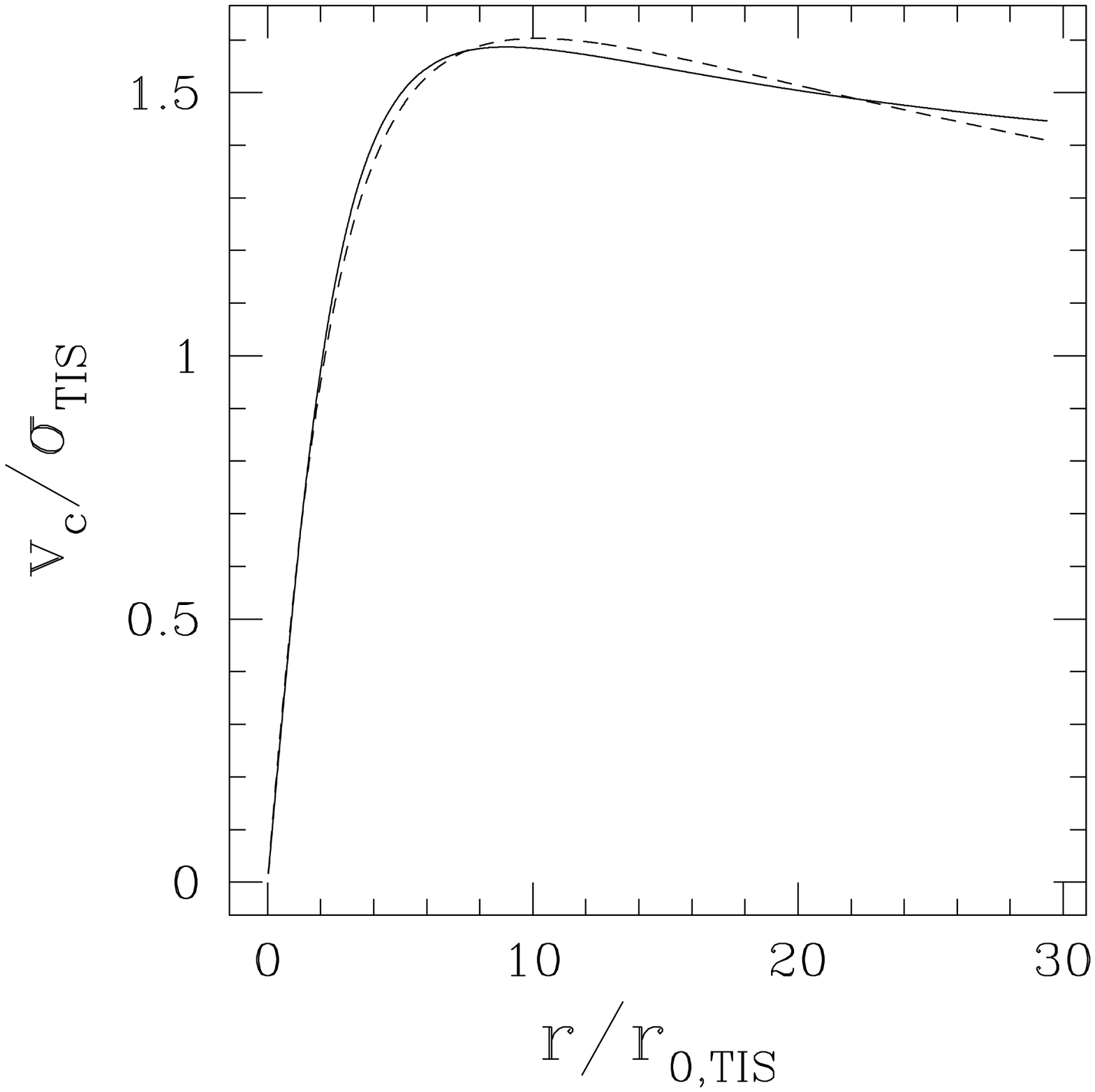,height=3.5in,width=4.5in}}

Fig. 3: Rotation Curve Fit. Best fit parameters:
$$\displaystyle{\frac{\rho_{0,Burkert}}{\rho_{0,TIS}}=1.216,\,
\frac{r_{c}}{r_{0,TIS}}=3.134}$$
Solid line = Best fit TIS; Dashed line = Burkert profile, where
$
\sigma_{TIS}^2=\langle v^2\rangle/3=k_BT/m.
$

Q: How well does this best fit TIS profile predict the $r_{max}$ and $v_{max}$?

A: $\displaystyle{\frac{r_{max,Burkert}}{r_{max,TIS}}
	=1.13,\,\frac{v_{max,Burkert}}{v_{max,TIS}}=1.01}$

(i.e. excellent agreement)

\pagebreak
\underline{The $v_{max}-r_{max}$ relation for dwarf and LSB galaxies.}

Q: Can the TIS halo model explain the observed correlation of $v_{max}$ and $r_{max}$ 
for dwarf spiral and LSB galaxies?

A: Yes, when the TIS halo model is combined with the Press-Schechter model which
predicts the typical collapse epoch for objects of a given mass (i.e. the mass of 
the $1\sigma$-fluctuations vs. $z_{coll}$). (See Fig. 4) For the three untilted
CDM models plotted, a cluster normalized Einstein-de Sitter model, and COBE-normalized
low-density models ($\Omega_0=0.3$ and $\lambda_0=0$ or 0.7), only the flat models
yield a reasonable agreement with the observed $v_{max}-r_{max}$ relation.
 
\centerline{\psfig{figure=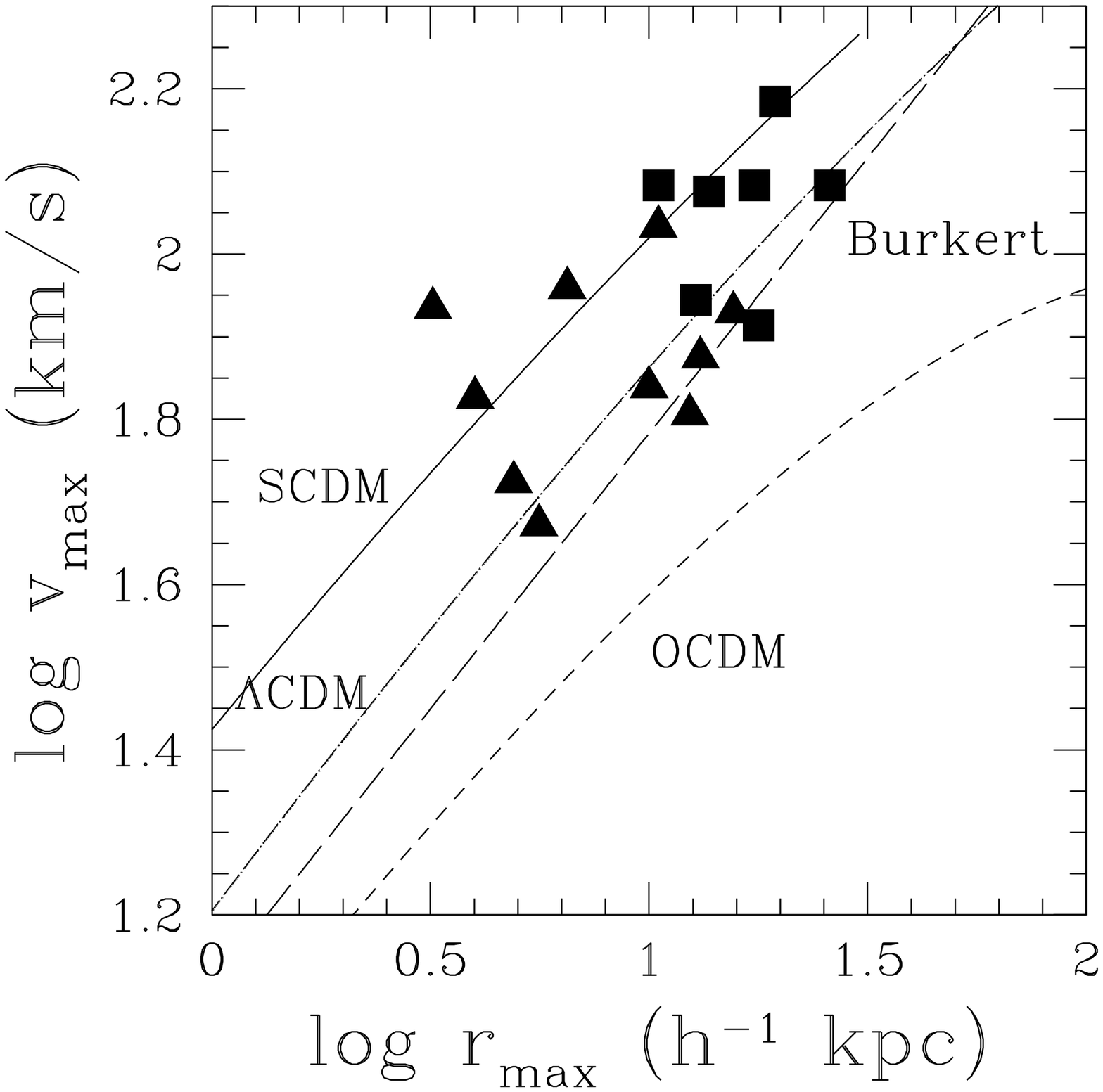,height=4.in,width=4.5in}}
Fig. 4: Dwarf galaxies (triangles) and LSB galaxies (squares) from Kravtsov et al. (1998);
Burkert: fit to data (Mori \& Burkert 2000); SCDM: $\Omega_0=1$, $\lambda_0=0$, 
$\sigma_{8h^{-1}}=0.5$ (cluster normalized); OCDM: $\Omega_0=0.3$, $\lambda_0=0$
 (COBE normalized); $\Lambda$CDM: $\Omega_0=0.3$, $\lambda_0=0.7$, (COBE normalized);
$h=0.65$ for all.

\pagebreak
\centerline{\underline{Galaxy Halo $M-\sigma_v$ Relation}}

Q: How well does our TIS halo model scaling relation predict the
velocity dispersion of galactic halos which form in the CDM model
according to N-body simulations? 

A: Antonuccio-Delogu, Becciani, \& Pagliaro (1999) used an N-body treecode 
at high-res ($256^3$ particles) to simulate galactic halos in the region
of a single and a double cluster. The agreement with the TIS model is good.

\centerline{\psfig{figure=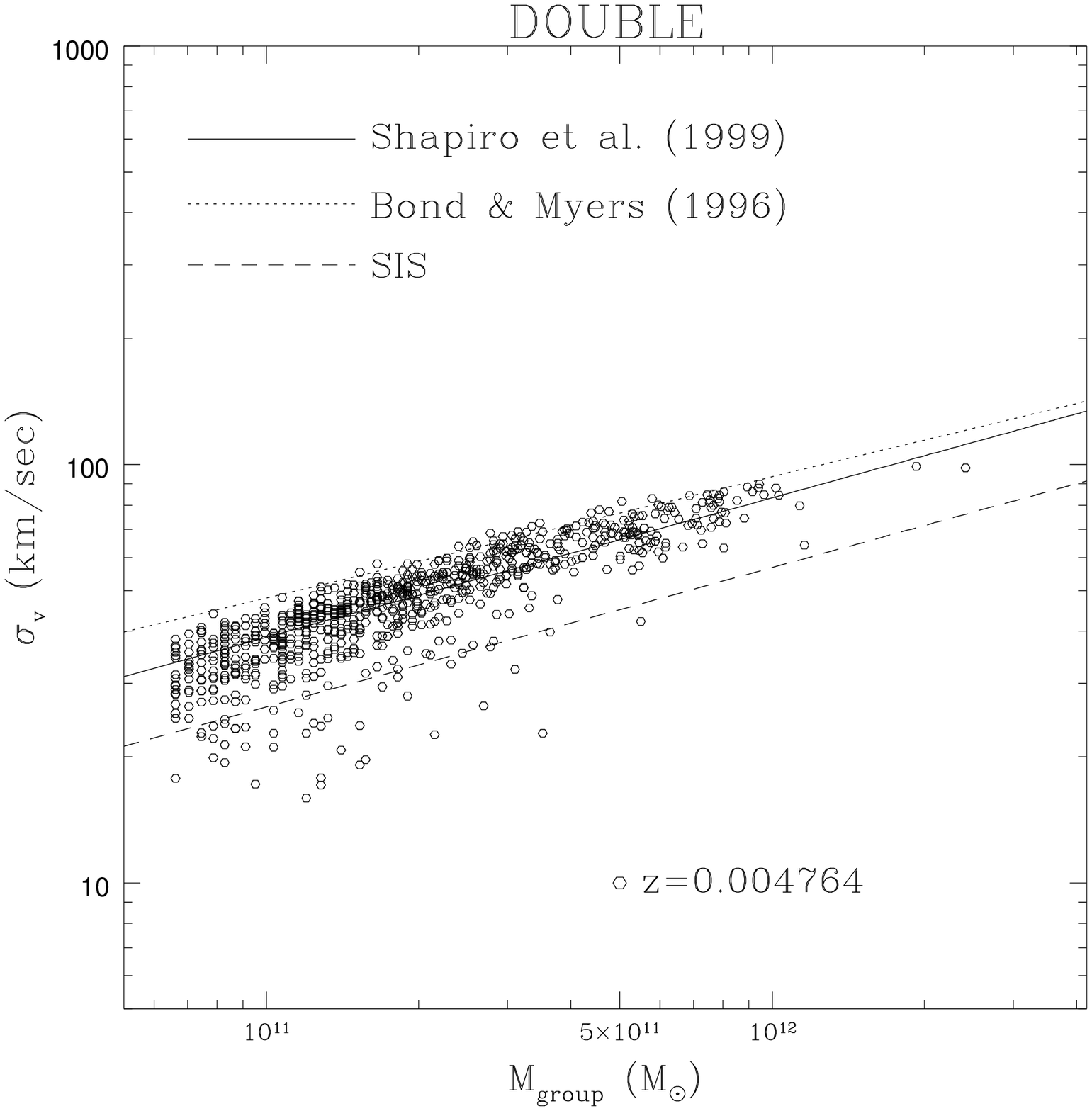,height=3.2in,width=4in}}
\centerline{\psfig{figure=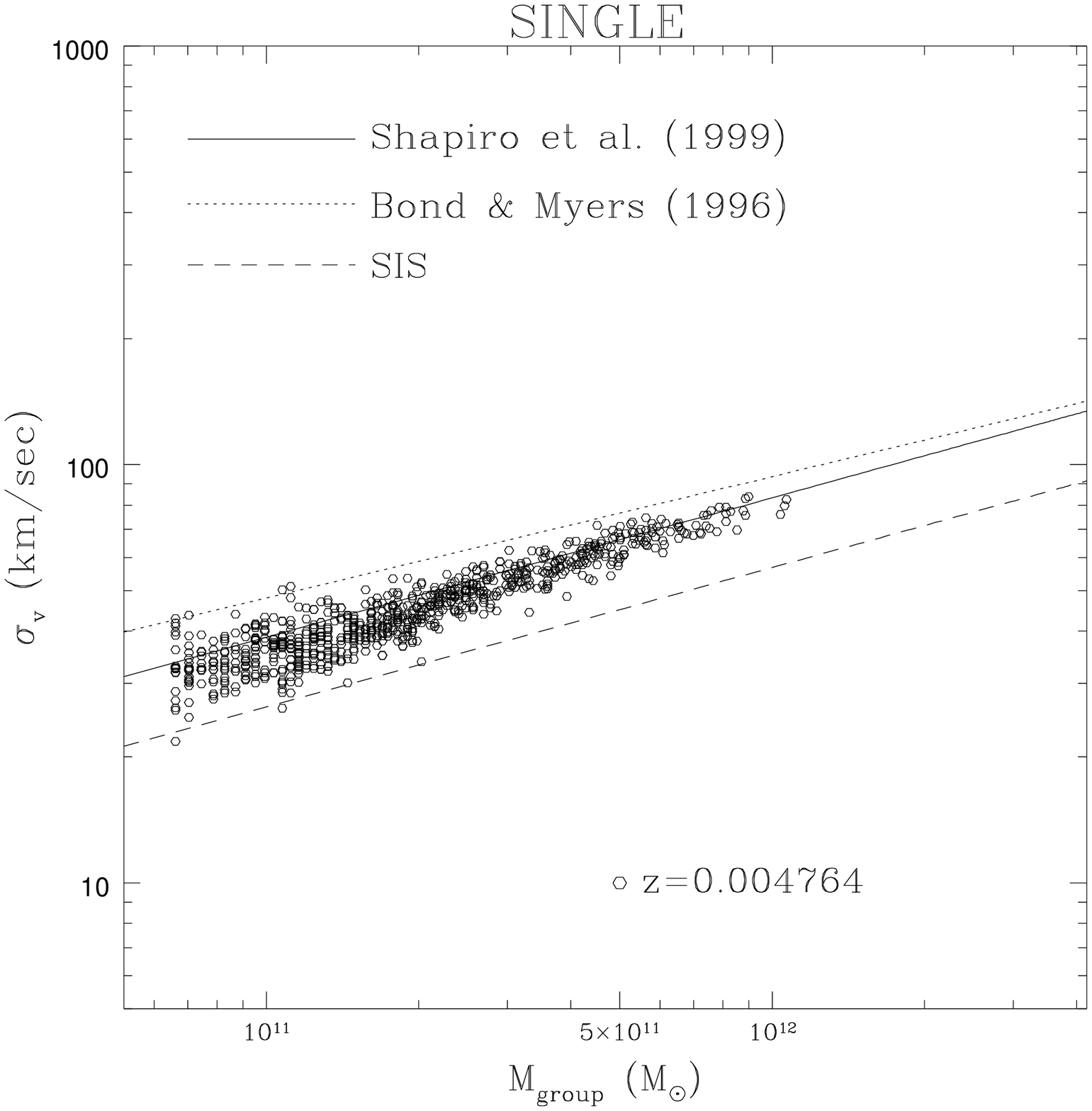,height=3.2in,width=4in}}

Fig. 6: Velocity dispersion vs. mass for galactic haloes in cluster 
regions: (upper panel) double cluster, (lower panel) single cluster.

\pagebreak

\centerline{\underline{X-Ray Cluster Scaling Relations}}

Q: How well does the TIS halo model predict the internal structure of X-ray
clusters found by gas-dynamical/N-body simulations of X-ray cluster 
formation in the CDM model?

A: As shown below and in Fig. 5, our TIS model predictions agree astonishingly well
with the mass-temperature and the radius-temperature virial relations and integrated 
mass profiles derived from numerical simulations by Evrard, Metzler and Navarro (1996).
Apparently, these simulation results are not sensitive to the discrepancy between our
prediction of finite density core and the N-body predictions of a density cusp for 
clusters in CDM.  
\begin{itemize}
\item Mass Profile -- Temperature Relation
$$
r_X\equiv r_{10}(X)\displaystyle{\left(\frac T{10\, {\rm keV}}\right)^{1/2}};\qquad
X\equiv\frac{\langle\rho(r)\rangle}{\rho_b}
$$
%\begin{minipage}{250mm}
%\begin{tabular}{@{}lccc}
%$X=\frac{\langle\rho\rangle}{\rho_b}$&$(r_{10})_{EMN}$&$(r_{10})_{TIS}$ & difference\\ \hline
%200..... &3.7 &3.759 &1.6\%\\
%250..... &3.37 &3.398 &0.8\%\\
%500..... &2.48 &2.479 &0.06\%\\
%1000...  &1.79 &1.796 &0.3\%\\
%2500...  &1.11 &1.137 &2.4\%\\ \hline
%\end{tabular}
%\end{minipage}
%\begin{itemize}
%\item 

\centerline{\psfig{figure=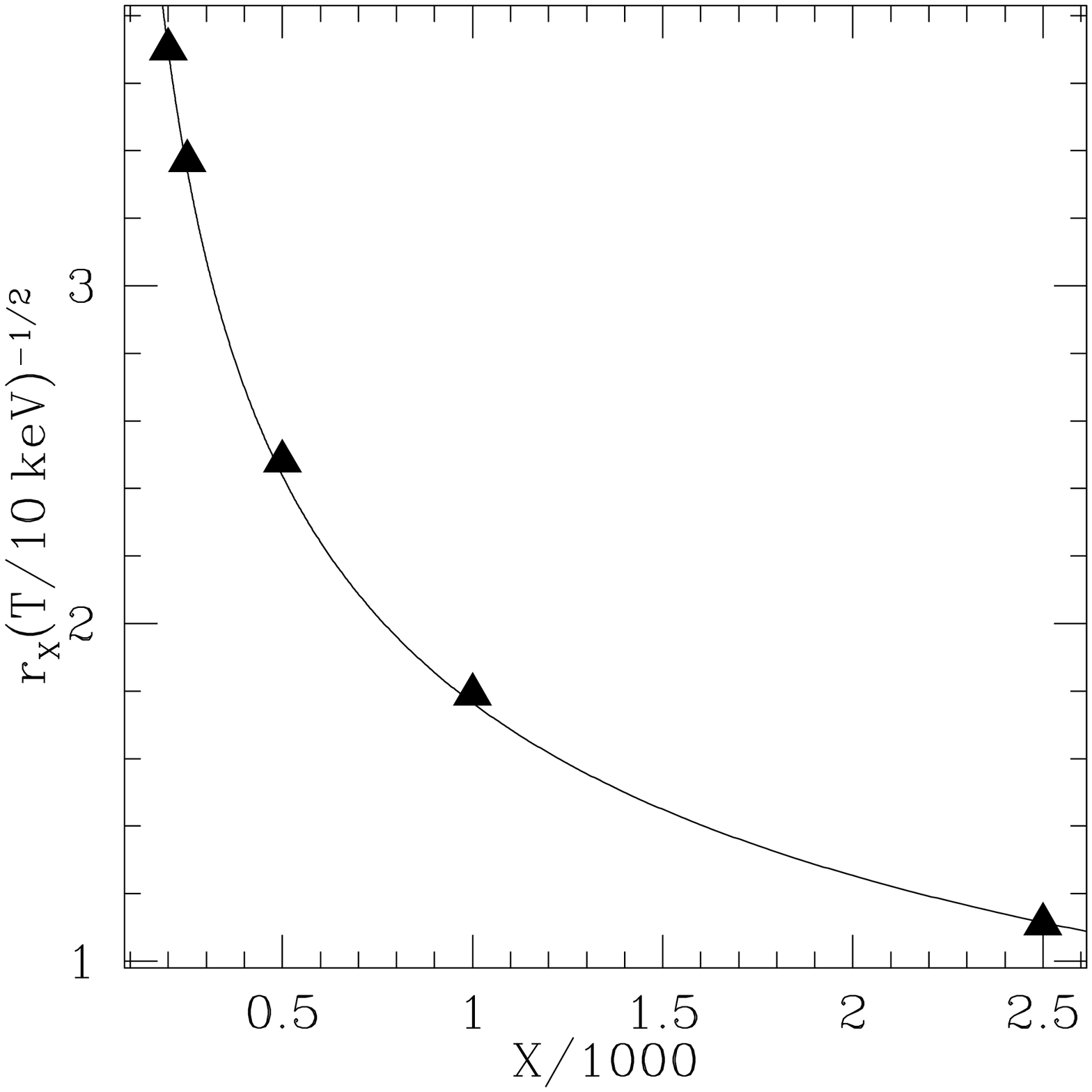,height=3.5in,width=5in}}

Fig.5 (triangles) fit to CDM simulation results by Evrard, Metzler and Navarro (1996);
(continuous line) TIS prediction.

\item Mass-Temperature and Radius-Temperature Virial Relations

-- \underline{Evrard, Metzler and Navarro (1996)}
\begin{eqnarray}
M_{500}
&=&\displaystyle{(1.11\pm 0.16)\times10^{15}\left(\frac T{10\, {\rm keV}}
	\right)^{3/2}} h^{-1}\, M_\odot,\nonumber\\
r_{500}
&=&\displaystyle{(1.24\pm 0.09)\left(\frac T{10\, {\rm keV}}\right)^{1/2}
		h^{-1}\, {\rm Mpc}}.\nonumber\\
M_{200}
&=&\displaystyle{1.45\times10^{15}\left(\frac T{10\, {\rm keV}}
	\right)^{3/2}} h^{-1}\, M_\odot,\nonumber\\
r_{200}
&=&\displaystyle{1.85\left(\frac T{10\, {\rm keV}}\right)^{1/2}
		h^{-1}\, {\rm Mpc}}.\nonumber
\end{eqnarray}

-- \underline{Our solution}
\begin{eqnarray}
M_{500}
&=&\displaystyle{1.11\times10^{15}\left(\frac T{10\, {\rm keV}}\right)^{3/2}}
		h^{-1}\, M_\odot,	\nonumber\\[5mm]
r_{500}
&=&\displaystyle{1.24\left(\frac T{10\, {\rm keV}}\right)^{1/2}
	\displaystyle{h^{-1}}{\rm Mpc}\,
			}.\nonumber\\[5 mm]
M_{200}
&=&\displaystyle{1.55\times10^{15}\left(\frac T{10\, {\rm keV}}
	\right)^{3/2}} h^{-1}\, M_\odot,\nonumber\\[5mm]
r_{200}
&=&\displaystyle{1.88\left(\frac T{10\, {\rm keV}}\right)^{1/2}
		h^{-1}\, {\rm Mpc}}.\nonumber
\end{eqnarray}
\end{itemize}

\pagebreak

\vspace{-1.3cm}
\centerline{\underline{$\beta$-fits to X-ray Cluster Brightness and }}
\centerline{\underline{ Density Profiles}}
\vspace{-1.3cm}
$$
\rho_{gas}=\frac{\rho_0}{\displaystyle{\left(1+r^2/r_c^2\right)^{3\beta/2}}},\,\,\,
I=\frac{I_0}{\displaystyle{\left(1+\theta^2/\theta_c^2\right)^{3\beta-1/2}}}
$$
Q: How well does the TIS model for the internal structure of X-ray clusters predict
the observed and simulated X-ray brightness profile of clusters?

A: It predicts gas density profiles and brightness profiles which are well-fit by
a $\beta-$profile, with $\beta$-values for the TIS $\beta$-fit which 
are close to 
those of simulated clusters in the CDM model, but somewhat larger than the 
conventional observational result that $\beta\approx2/3$. However, recent X-ray results
suggest that the true $\beta$-values are larger than 2/3 when measurements at larger 
radii are used and when central cooling flows are excluded from the fit.

{\tiny
\begin{minipage}{250mm}
\begin{tabular}{@{}lc}
{\bf Brightness profile observations} &$\beta$\\\hline
Jones and Foreman (1999)& 0.4-0.8, ave. 0.6\\[3mm]
Jones and Foreman (1992)& $\sim 2/3$\\[3mm]
Balland and Blanchard (1997)& 0.57 (Perseus)\\
	&	0.75 (Coma)\\[3mm]
Durret et al. (2000) & 0.53 (incl. cooling flow)\\
	&	0.82 (excl. cooling flow)\\ 
Vikhlinin, Forman, \& Jones (1999)&	0.7-0.8\\
(fit by Henry 2000)&\\
\hline
TIS $\beta$-fit ($r_c/r_{0,TIS}=2.639$)& 0.904\\\hline\\[3mm]
{\bf Gas density profile simulations}&$\beta$\\\hline
Metzler and Evrard (1997) & 0.826 (DM)\\
&	 0.870 (gas)\\[3mm]
Eke, Navarro, and Frenk (1998) & 0.82\\[3mm]
Lewis et al. (1999) (adiabatic) &	$\sim 1$\\[3mm]
Takizawa and Mineshige (1998) & $\sim 0.9$ \\[3mm]
Navarro, Frenk, and White (1995)& 0.8\\[3mm]\hline
TIS $\beta$-fit ($r_c/r_{0,TIS}=2.416$) & 0.846\\\hline
\end{tabular}
\end{minipage}}

\pagebreak

\centerline{\underline{X-ray Cluster Gas Entropy}}

Q: Can the TIS model for the internal structure of X-ray clusters explain the
observed correlation between the gas entropy near the cluster center and cluster
virial temperature?

A: Yes, but only for high-T clusters (i.e. $T>$ few keV) for which energy release
feedback effects were probably not big enough to alter the entropy of the equilibrium
halo. (See Fig. 7)
$$
S=T/n_e^{2/3}, \mbox{at $r=0.1r_{vir}$}
$$

\vspace{-1cm}
\centerline{\psfig{figure=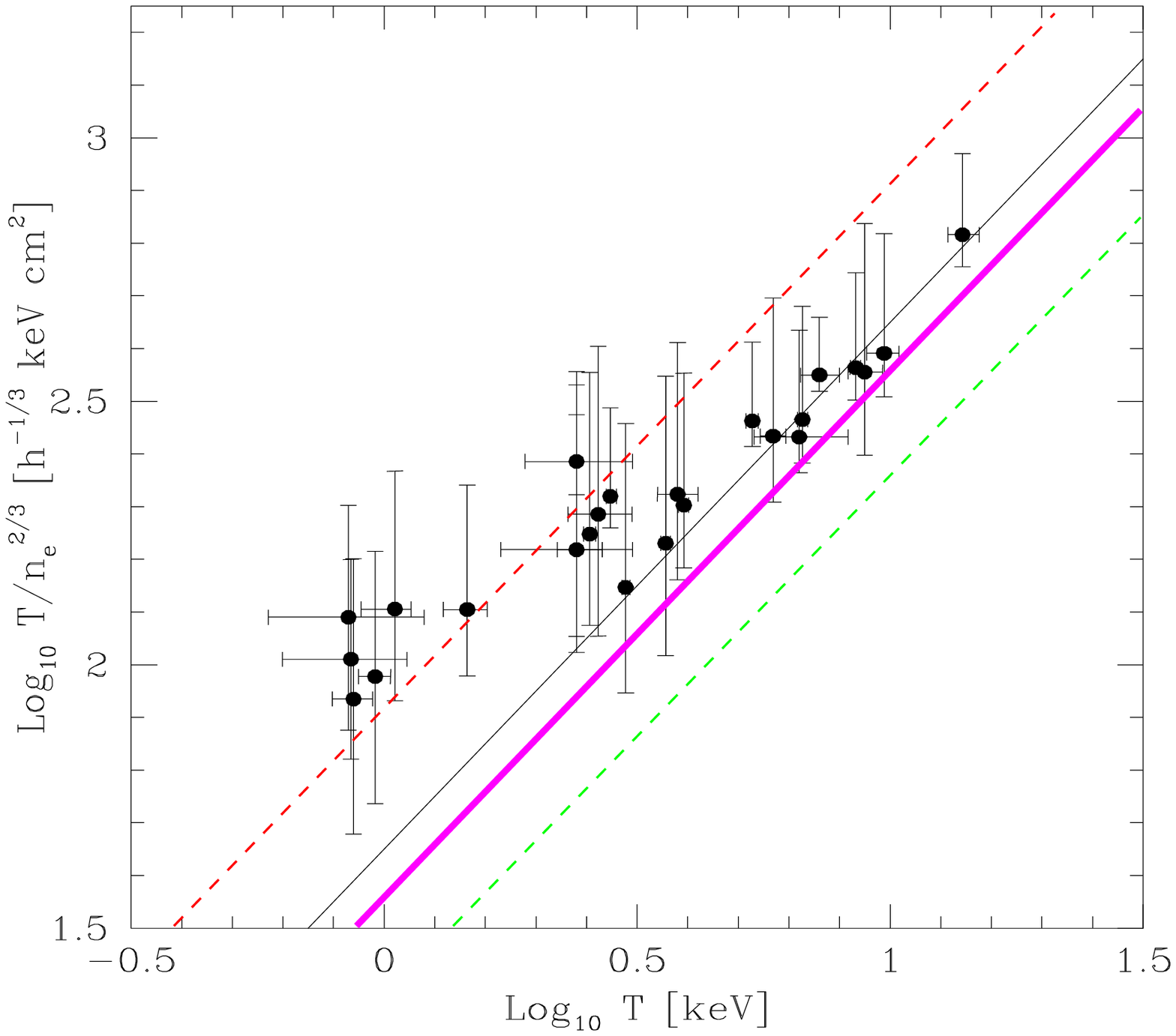,height=5in,width=5in}}
Fig. 7: Cluster entropy vs. temperature. Data: Ponman, Cannon, and Navarro (1999) 
Nature, 397; Error bars: T -- 90\% confidence level, entropy -- span of variation 
from $r=0.05r_{vir}$ to $r=0.2r_{vir}$. Our solution: thick line = S at $r=0.1r_{vir}$,
dashed lines = S at $r=0.05r_{vir}$ (lower), and $0.2r_{vir}$ (upper).
\pagebreak

\centerline{\underline{Cluster Mass Profiles Deduced from
Strong Gravitational Lensing}} 

Q: Can the TIS halo model explain the mass profile with a finite density core 
measured by Tyson, Kochanski, and Dell'Antonio (1998) for cluster CL 0024+1654
at $z=0.39$ using the strong gravitational lensing of background galaxies by the 
cluster to infer the cluster mass distribution?

A: Yes, the TIS model not only provides a good fit to the shape of the projected 
surface mass density distribution of this cluster (see Fig. 8) within the arcs, but
when we match the central value as well as the shape, our model predicts the overall 
mass, and a cluster velocity dispersion in close agreement with the value 
$\sigma_v=1200$ km/s measured by Dressler and Gunn (1992).

\centerline{\psfig{figure=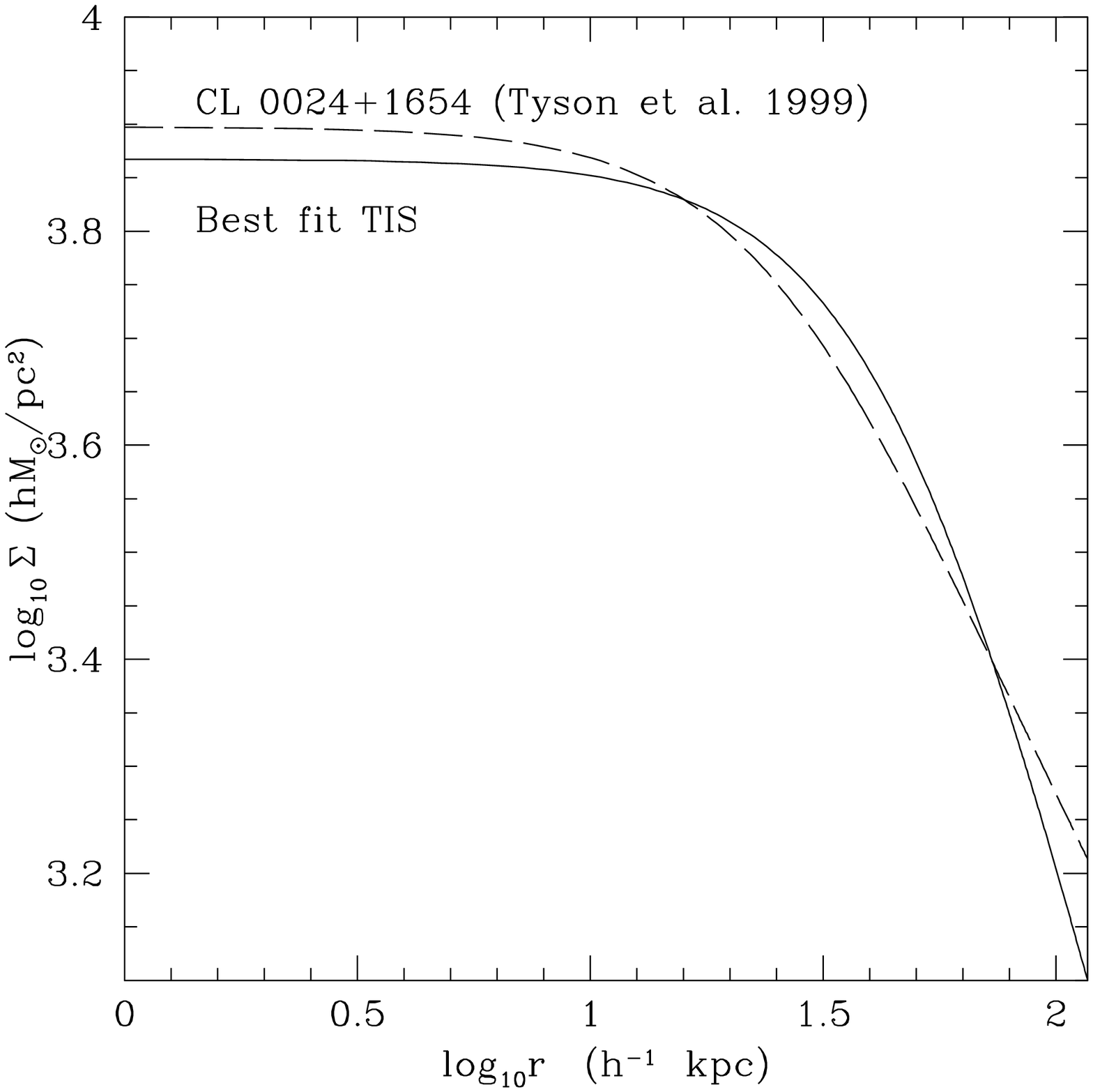,height=4.5in,width=4.5in}}

Fig. 8: Projected surface density of cluster CL 0024+1654 inferred from lensing 
measurements, together with the best fit TIS model.
\pagebreak

\centerline{\underline{Summary}}
{\footnotesize 
\begin{itemize}
\item TIS profile fits dwarf galaxy rotation curves; combined with Press-Schechter
formalism matches results for observed $v_{max}-r_{max}$ relation for dwarf 
galaxies
\item Predicted mass-velocity dispersion relation 
agrees with high resolution N-body simulations of galactic halo formation by
Antonuccio-Delogu et al. (1999)
\item Predicted mass-radius-temperature scaling relations match
simulation results from X-ray clusters in CDM model
\begin{itemize}
\item Our solution {\bf derives} empirical fitting formulae of 
Evrard, Metzler and Navarro (1996)
\item Agrees well with X-ray cluster observations at $z=0$
\end{itemize}
\item Fits high temperature X-ray cluster entropy data
\item X-ray brightness profile is predicted to match $\beta$-fit with 
$\beta\approx0.9$, larger than typically observed, but similar to results of
gas-dynamical/N-body simulations of X-ray clusters in CDM model
\item Fits the cluster mass profile with finite core derived from 
strong gravitational lensing data of Tyson et al. (1999) on CL 0024+1654
\item Predicted mass profile is close to NFW profile for low values of 
concentration parameter, outside the core
\end{itemize}}

%\acknowledgements 
{\tiny This work was supported in part by NASA Grants NAG5-2785, NAG5-7363, and
NAG5-7821, NSF Grant ASC-9504046, and Texas Advanced Research Program
Grant 3658-0624-1999, and benefitted from PRS' participation at the 
Aspen Center for Physics in summer 1998 and 1999.}

\pagebreak
\centerline{\underline{References}}
\tiny
Antonuccio-Delogu, Becciani, \& Pagliaro (1999) To appear in the proceedings of 
the Marseille IGRAP99 conference "Clustering at High Redshift", astro-ph/9910266

Balland C., and Blanchard A. (1997), ApJ, 487, 33

Bertschinger E. 1985, ApJS, {  58}, 39

Burkert A. 1995, ApJ, 447, L25

Dressler and Gunn 1992, ApJS, 78, 1

Durret et. al. 2000, astro-ph/0002519

Eke V.R.,  Navarro J.F., and  Frenk C.S. 1998, ApJ {  503}, 569

Evrard A.E., Metzler C.A., and Navarro J.F. 1996, ApJ, {  469}, 494

Henry P. 2000, astro-ph/0002365

Iliev \& Shapiro 2000, in preparation

Jones C., and Forman W., 1992, in Clusters and Superclusters of 
Galaxies, ed. A. C. Fabian
Kluwer Academic Publishers, pp 49-70

Jones C., and Forman W. 1999, ApJ, 511, 65

Kravtsov A. V., Klypin A. A., Bullock J. S., and Primack J. R. 1998, 
	ApJ, {  502}, 48

Lewis, G. F., Babul A., Katz N., Quinn T., Hernquist L., Weinberg D.H., astro-ph/9907097

Metzler C.A., and Evrard A.E., astro-ph/9710324

Mori M., \& Burkert A. 2000, astro-ph/0001422

Navarro J.F., Frenk C.S., and White S.D.M. 1995, MNRAS, 275, 720 

Navarro J.F., Frenk C.S., and White S.D.M. 1996, ApJ, {  462}, 563

Navarro J.F., Frenk C.S., and White S.D.M. 1997, ApJ, {  490}, 493

Ponman T.J., Cannon D.B., and Navarro J.F. 1999, Nature, 397, 135

Shapiro, Iliev, and Raga 1999, MNRAS, 307, 203

Takizawa M., Mineshige S. 1998, ApJ, 499, 82 

Tyson J.A., Kochanski, G.P., and Dell'Antonio I.P. 1998, ApJ, 
	{  498}, L107

Vikhlinin A., Forman W., \& Jones C. 1999, astro-ph/9905200

}
%\end{thebibliography}

\end{document}